\documentclass{article}
\usepackage{spconf,amsmath,graphicx,booktabs}
\usepackage{multirow}
\usepackage{color}
\usepackage{algorithm}
\usepackage[noend]{algpseudocode}

\DeclareMathOperator\tr{Tr}

\newcommand{\MB}[1]{\mbox{\boldmath{$#1$}}}

\newcommand{\loss}{{\cal L}}
\newcommand{\U}{\MB{U}}
\newcommand{\V}{\MB{V}}
\newcommand{\W}{\MB{W}}
\newcommand{\G}{\MB{G}}
\newcommand{\hatW}{\hat{\W}}
\newcommand{\deltaU}{\delta\U}
\newcommand{\deltaV}{\delta\V}
\newcommand{\deltaW}{\delta\W}
\newcommand{\deltaL}{\delta\loss}
\newcommand{\tildeW}{\tilde{\W}}
\newcommand{\grad}[1]{\nabla \loss(#1)}

\definecolor{gray}{rgb}{0.3,0.3,0.3}
\definecolor{darkgreen}{rgb}{0.1,0.7,0.1}
\definecolor{darkpurple}{rgb}{0.7,0.1,0.7}
\definecolor{darkbrown}{rgb}{0.6,0.4,0.1}

\newcommand{\comment}[1]{\State {\em \color{gray} #1}}

\title{LOW-RANK GRADIENT APPROXIMATION FOR MEMORY-EFFICIENT ON-DEVICE TRAINING OF DEEP NEURAL NETWORK}
\name{Mary Gooneratne\sthanks{Work performed as an intern at Google.}, Khe Chai Sim, Petr Zadrazil, Andreas Kabel, Fran{\c{c}}oise Beaufays, Giovanni Motta}
\address{Google, USA\\
\texttt{mary.gooneratne@duke.edu} \\
\texttt{\{khechai,binus,aka,fsb,giovannimotta\}@google.com}
}

\begin{document}
\ninept
\maketitle
\begin{abstract}
Training machine learning models on mobile devices has the potential of improving both privacy and accuracy of the models. However, one of the major obstacles to achieving this goal is the memory limitation of mobile devices.
Reducing training memory enables models with high-dimensional weight matrices, like automatic speech recognition (ASR) models, to be trained on-device. 
In this paper, we propose approximating the gradient matrices of deep neural networks using a low-rank parameterization as an avenue to save training memory. 
The low-rank gradient approximation enables more advanced, memory-intensive optimization techniques to be run on device.
Our experimental results show that we can reduce the training memory by about 33.0\% for Adam optimization. It uses comparable memory to momentum optimization and achieves a 4.5\% relative lower word error rate on an ASR personalization task. 
\end{abstract}
\begin{keywords}
on-device learning, low-rank gradient, memory reduction
 \end{keywords}
\section{Introduction}
\label{sec:intro}

State-of-the-art speech-recognition models are based on deep neural networks~\cite{hinton2012deep} with weight matrices of dimensions in the order of thousands. We have shown that such models can be deployed offline on mobile devices~\cite{he2019streaming}. 
Decentralizing the training of these models to be on-device can improve personalization and security. However, the advanced optimization techniques used to train models require additional memory proportional in size to the model parameters. Therefore, one of the major obstacles to achieving high-accuracy on-device models is the memory limitation of devices for training. Previous explorations to reduce training memory included training only on part of the model and/or splitting the gradient computation into multiple steps~\cite{sim19interspeech,sim19asru}.

In this paper, we propose using low-rank gradient approximation to reduce the training memory needed for advanced optimization techniques. Note that there are already methods in the literature that use low-rank structure to reduce model size~\cite{xue2014singular,zhao2016low,samarakoon2016factorized}. This proposal does not apply the low-rank approximation to the weight matrices but rather to the gradients, as to retain the full modeling power of the model.
The approach we take is less computationally expensive than Singular Value Decomposition (SVD)~\cite{nash1990singular} and, again, does not constrain the model parameters by using the approximation exclusively as a vehicle for gradient computation. 

The remainder of the paper is organized as follows.
Section~\ref{sec:optimization} describes several optimization techniques and their associated training memory. 
Section~\ref{sec:low_rank_gradient} presents the proposed low-rank gradient optimization method.
Section~\ref{sec:discussions} analyses the effects of low-rank gradient approximation on training speed and convergence.
Section~\ref{sec:experiments} presents the experimental results on a personalization tasks for on-device speech recognition.

\section{Parameter Optimization}
\label{sec:optimization}

Deep neural networks are optimized by minimizing a loss function, which is a nonlinear function of the model parameters. This is done iteratively by updating the parameters in a direction that reduces the loss function. Let's denote the loss by $\loss(\W)$, where $\W$ is the weight matrix to be updated.
The update formula can be computed as:
\begin{equation}
\W_{t+1} \leftarrow \W_{t} + \deltaW_{t} 
\end{equation}
where $\W_t$ and $\deltaW_t$ are the weight matrix and the corresponding update at training step $t$.
For gradient descent optimization,
the update direction is given by the negative of the gradient of the loss with respect to the weight matrix:
\begin{equation}
\deltaW_t = - \lambda \grad{\W_t} \label{eqn:gradient_descent}
\end{equation}
where $\lambda$ is the learning rate and $\grad{\W_t}$ is the gradient at $\W_t$,
which can be computed using error back propagation~\cite{mcclelland1987parallel}.
There are more advanced optimization techniques that compute a better update direction to improve training convergence. For example, the update direction for momentum optimization~\cite{sutskever2013importance} is recursively computed as follows:
\begin{equation}
\deltaW_t = \mu \deltaW_{t - 1} - \lambda \grad{\W_t} \label{eqn:momentum_recursion}
\end{equation}
where $\mu \in [0, 1]$ is the momentum coefficient. Additional memory is required to save $\deltaW_{t-1}$ (same size as $\W$) for the subsequent training step (doubling the model size).
For Adam optimization~\cite{kingma2015adam}, the update direction is given by:
\begin{equation}
\deltaW_t = - \lambda \frac{\sqrt{1 - \beta_1^t}}{1 - \beta_2^t} \MB{M}_t \oslash \left(\sqrt{\MB{V}_t} - \epsilon\right)
\end{equation}
where $\beta_1$, $\beta_2$ and $\epsilon$ are scalar parameters.
The first and second momentum terms, $\MB{M}_t$ and $\MB{V}_t$, are computed recursively as:
\begin{eqnarray}
    \MB{M}_t &=& \beta_1 \MB{M}_{t-1} + (1 - \beta_1) \grad{\W_t}  \label{eqn:adam_recursion1} \\
    \MB{V}_t &=& \beta_2 \MB{V}_{t-1} + (1 - \beta_2) \grad{\W_t} \odot \grad{\W_t} \label{eqn:adam_recursion2} 
\end{eqnarray}
The symbols $\odot$ and $\oslash$ denote the element-wise multiplication and division operators.
Two additional terms are introduced, which results in a memory requirement that is 3 times the size of the original model.

\section{Low-rank Gradient Approximation}
\label{sec:low_rank_gradient}

As shown in the previous section, advanced optimization techniques require more memory to store additional terms. To reduce the total amount of memory required, we propose using low-rank gradient approximation.
Although low-rank approximation can be achieved using Singular Value Decomposition (SVD)~\cite{nash1990singular}, applying SVD to the gradients for each training step is computationally expensive.
Instead, we propose to re-parameterize the weight matrix into two parts:   
\begin{equation}
\W = \tildeW + \U \V^\top
\label{eqn:low_rank_reparam}
\end{equation}
where $\tildeW$ is an unconstrained matrix with the same size as $\W$, and $\U\V^\top$ is a low-rank matrix of rank $R$.
If $\W$ is a $M \times N$ matrix, then $\U$ and $\V$ are matrices of sizes
$M \times R$ and $N \times R$, respectively ($M \gg R$, $N \gg R$).  
With the re-parameterization in Eq.~\ref{eqn:low_rank_reparam}, we can reduce training memory by keeping $\tildeW$ fixed and updating only $\U$ and $\V$. However, this leads to a {\em low-rank model}, where the model parameter space is constrained to be low rank.
In order to keep the model parameters unconstrained, we treat $\tildeW$ as the actual model parameters, and use $\U$ and $\V$ only for the purpose of gradient computation.
Therefore, the update of $\W$ is constrained to be low-rank by those of $\U$ and $\V$ such that the effective gradient of $\W$ is given by:
\begin{equation}
    \hat{\grad{\W}} \approx \U \grad{\V}^\top + \grad{\U}\V^\top \label{eqn:grad_w_approx}
\end{equation}
The gradients of $\U$ and $\V$ can be computed from the gradient of $\W$ as follows:
\begin{eqnarray}
\grad{\U} &=& \grad{\W} \V \label{eqn:dldu} \\
\grad{\V} &=& \grad{\W}^\top \U \label{eqn:dldv} 
\end{eqnarray}
By substituting Eq.~\ref{eqn:dldu} and~\ref{eqn:dldv} into Eq.~\ref{eqn:grad_w_approx}
we can rewrite the effective gradient of $\W$ as:
\begin{equation}
    \hat{\grad{\W}} \approx \U\U^\top \grad{\W} + \grad{\W} \V\V^\top \label{eqn:grad_w_proj}
\end{equation}
where $\U\U^\top$ and $\V\V^\top$ are the low-rank projections of the rows and columns of $\W$, respectively.

It is useful to note that we can compute $\grad{\W}$ from the original model and then compute the gradients for $\U$ and $\V$ using Eq.~\ref{eqn:dldu} and~\ref{eqn:dldv}.
The re-parameterization in Eq.~\ref{eqn:low_rank_reparam} needs not be explicitly applied to the model
({\em i.e.} no need to modify the model computational graph). Instead, it can be applied by post-processing the gradient. 
This makes it easy to apply low-rank gradient training to existing models.

For the case of gradient descent, the update direction is given by the negative of the gradient scaled by the learning rate (Eq.~\ref{eqn:gradient_descent}). From Eq.~\ref{eqn:grad_w_approx}, we get
\begin{equation}
    \deltaW \approx \U \deltaV^\top + \deltaU\V^\top \label{eqn:delta_w}
\end{equation}
and the corresponding change in the loss function (ignoring the higher-order terms):
\begin{eqnarray}
    \deltaL &\approx& - \lambda \tr \left( \U\U^\top \G\G^\top + \V\V^\top\G^\top\G \right) \nonumber\\
    &=& - \lambda \tr \left( \U^\top \G\G^\top\U + \V^\top\G^\top\G\V \right) \label{eqn:delta_loss}
\end{eqnarray}
where we define $\G = \grad{\W}$ for clarity.
With the cyclic invariance of the trace, we can express the terms inside the trace as
positive semi-definite $R \times R$ matrices.
This will result in a non-positive change to the loss ($\deltaL \le 0$), as $\lambda > 0$ and the trace of a positive semi-definite matrix is non-negative.
Note that in the unrestricted case (without low-rank projection), the change in loss is given by:
\begin{equation}
    \deltaL = - \lambda \frac{1}{2} \tr \left( \G\G^\top + \G^\top\G \right) \label{eqn:delta_loss_full}
\end{equation}
In the special case where $\U$ and $\V$ are orthogonal matrices, the projection matrices $\MB{P}_u = \U\U^\top$ and $\MB{P}_v = \V\V^\top$ are diagonal matrices with elements 1 or 0.
In fact, they are rank-$R$ matrices with exactly $R$ entries of 1's on the leading diagonal. A smaller $R$ will result in a smaller trace term in Eq.~\ref{eqn:delta_loss}, and therefore a smaller reduction in loss.
As a result, we expect low-rank approximation to slow down training convergence.

\subsection{Random Gradient Projection}

From Eq.~\ref{eqn:dldu} and~\ref{eqn:dldv}, if $\U$ and $\V$ are zero matrices, their gradients will also be zero. Therefore, we need to assign non-zero values to $\U$ and $\V$ at each training step so that they can be updated. 
Ideally, we want to choose $\U$ and $\V$ to maximize the magnitude of $\deltaL$ in Eq.~\ref{eqn:delta_loss} using SVD. However, this will be computationally expensive.
Instead, we assign them with random values.
It can be shown that by drawing random values from a zero-mean normal distribution with standard deviations of $\frac{1}{\sqrt{N}}$ and $\frac{1}{\sqrt{M}}$ for $\U$ and $\V$, respectively,
$\U^\top\U$ and $\U^\top\U$ are close to an $R \times R$ identity matrix ($\U$ and $\V$ are approximately orthogonal).
Furthermore, by comparing between Eq.~\ref{eqn:delta_loss} and~\ref{eqn:delta_loss_full}, it is desirable to choose $\U$ and $\V$ such that the eigenvalues of $\U\U^\top$ and $\V\V^\top$ are close to $\frac{1}{2}$.
This can be accomplished by drawing the values of $\U$ and $\V$ from ${\cal N}\left(\MB{0}^{M \times R}, \frac{1}{\sqrt{2M}}\MB{I}^{M \times R}\right)$ and ${\cal N}\left(\MB{0}^{N \times R}, \frac{1}{\sqrt{2N}}\MB{I}^{N \times R}\right)$, respectively.

\subsection{Implementations}

The low-rank gradient approximation method described above can be implemented by adding new variables $\U$ and $\V$ for each weight matrix in the model.
Note that constraining the gradient of $\W$ to be low-rank (using Eq.~\ref{eqn:grad_w_proj}) does not necessarily yield a low-rank momentum term ({\em e.g} Eq.~\ref{eqn:adam_recursion2}).
Instead, it is easier to keep track of the momentum terms by updating $\U$ and $\V$ separately and use the updated $\U$ and $\V$ to update $\W$ using Eq.~\ref{eqn:low_rank_reparam}.
If $\U$ and $\V$ are updated by $\deltaU$ and $\deltaV$ respectively, the effective update of $\W$ is given by
\begin{align}
\W_{t+1} &\leftarrow \tildeW + \left( \U_t + \deltaU_t \right) \left( \V_t + \deltaV_t \right)^\top \label{eqn:low_rank_update} \\
&= \W_t + \underbrace{\U_t\deltaV_t^\top + \deltaU_t\V_t^\top + \deltaU_t\deltaV_t^\top}_{\deltaW_t} \label{eqn:low_rank_delta_w} 
\end{align}
Note the additional second-order term ($\deltaU\deltaV^\top$) in Eq.~\ref{eqn:low_rank_delta_w} compared to Eq.~\ref{eqn:delta_w}.
This way, we are able to combine low-rank gradient approximation with existing advanced optimization techniques.
In fact, $\deltaW$ in Eq.~\ref{eqn:low_rank_delta_w} can be rewritten as:
\begin{equation}
    \deltaW = \U_{t+1}\V_{t+1}^\top - \U_t \V_t^\top \label{eqn:update_w}
\end{equation}
That is, the update direction is given by the difference between the new and old low-rank matrix, $\U\V^\top$.

\begin{algorithm}[t]
\caption{Low-rank Gradient Approximation Algorithm}\label{algo:low_rank}
\begin{algorithmic}[1]
\Procedure{LowRankUpate}{$\W_t, \grad{\W_t}, R$}
    \comment{\# Randomize $\U$ and $\V$.}
	\State $\U \sim {\cal N}\left(\MB{0}^{M \times R}, \frac{1}{\sqrt{2M}}\MB{I}^{M \times R}\right)$.  \label{algo:low_rank:randomize_u}
	\State $\V \sim {\cal N}\left(\MB{0}^{N \times R}, \frac{1}{\sqrt{2N}}\MB{I}^{N \times R}\right)$.  \label{algo:low_rank:randomize_v}
    \comment{\# Compute gradients.}
	\State $\grad{\U_t} \leftarrow \grad{\W_t} \V_t$ (using Eq.~\ref{eqn:dldu}) \label{algo:low_rank:grad_u}
	\State $\grad{\V_t} \leftarrow \grad{\W_t}^\top \U_t$ (using Eq.~\ref{eqn:dldv}) \label{algo:low_rank:grad_v}
    \comment{\# Update $\U$ and $\V$.}
	\State $\U_{t+1} \leftarrow \U_t + \deltaU_t$ \label{eqn:update_u}
	\State $\V_{t+1} \leftarrow \V_t + \deltaV_t$ \label{eqn:update_v}
    \comment{\# Update $\W$}
	\State $\W_{t+1} \leftarrow \W_{t} + \U_{t+1}\V_{t+1}^\top - \U_{t}\V_{t}^\top$
	(using Eq.~\ref{eqn:update_w})
	\label{algo:low_rank:add_uv}
\EndProcedure
\end{algorithmic}
\end{algorithm}
The algorithm for computing the low-rank gradients is shown in Algorithm~\ref{algo:low_rank}.
For each training step, we first assign random values to $\U$ and $\V$ (lines~\ref{algo:low_rank:randomize_u} and~\ref{algo:low_rank:randomize_v}).
Next, we compute the gradients for $\U$ and $\V$ (lines~\ref{algo:low_rank:grad_u} and~\ref{algo:low_rank:grad_v}) and updates $\U$ and $\V$ using standard optimization techniques,
such as gradient descent, momentum, and Adam (lines~\ref{eqn:update_u} or~\ref{eqn:update_v}).
Finally, in line~\ref{algo:low_rank:add_uv}, we update $\W$ using Eq.~\ref{eqn:update_w}.

\section{Analysis}
\label{sec:discussions}

We set up a simple problem to analyze and understand the behaviour of the proposed low-rank gradient method. The goal is to learn a matrix $\W$ to match a target matrix, $\hatW$.
The following mean squared error loss function is used:
\begin{equation}
    {\cal L}(\W) = \frac{1}{D^2} \sum_{i,j} \left( \exp(w_{i,j}) - \exp(\hat{w}_{i,j}) \right)^2
\end{equation}
where $\W$ and $\hatW$ are matrices of size $D \times D$. 
The $(i,j)$-th element of $\W$ and $\hatW$ are denoted by $w_{i,j}$ and $\hat{w}_{i,j}$, respectively.
We use $\exp(\cdot)$ to introduce non-linearity to the function.
We compared using different optimization methods and low-rank projection methods. {\tt none} means that there is no low-rank gradient approximation, {\tt random} refers to the case where $\U$ and $\V$ are randomly set per training step and {\tt svd} means that $\U$ and $\V$ are estimated by approximating the gradient of $\W$ using SVD.
We performed 50,000 training steps for each configuration.
\begin{table}[t]
    \centering
    \caption{Comparing loss and training time after 50,000 training steps for different optimization methods ($D=100$, $R=5$).}
    \label{tab:loss_training_speed}
    \begin{tabular}{c|c|c|c}
        \toprule
        Optimization & Projection &
        \multirow{2}*{~~Loss~~}   & 
        Training time \\
        Method  &   Method  &           & (seconds)   \\\midrule
        \multirow{3}*{Gradient Descent}
        &   none      & 0.00510        & 22.4 \\
        &   random    & 1.73146        & 24.0 \\
        &   svd       & 0.44033        & 329.2 \\\hline
        \multirow{3}*{Momentum}
        &   none      & 0.00059        & 20.2 \\
        &   random    & 1.72933        & 29.5 \\
        &   svd       & 0.31221        & 347.2 \\\hline
        \multirow{3}*{Adam}
        &   none      & 0.00026        & 28.0 \\
        &   random    & 0.00008        & 33.2 \\
        &   svd       & 0.00028        & 317.6 \\
        \bottomrule
    \end{tabular}
\end{table}
Table~\ref{tab:loss_training_speed} shows the loss and training time after 50,000 training steps.
In general, {\tt svd} approximation yields a much better loss value after 50,000 training steps across different optimization methods, except for Adam optimization where all methods converged to a loss value of less than $10^{-3}$. On the other hand, the {\tt random} method is only slightly slower than the standard method while the {\tt svd} method takes an order magnitude longer time to train (due to the need to computed SVD every training step).

\section{Experimental Results}
\label{sec:experiments}

We collected a dataset we called {\em Wiki-Names}~\cite{sim19asru} to evaluate the performance of speech personalization algorithms. The text prompts are sentences extracted from English Wikipedia pages that contain repeated occurrences of politician and artist names that are unique and difficult to recognize (we selected them by synthesizing speech for these names and verifying that our baseline recognizer recognizes them incorrectly).

The dataset aggregates speech data from 100 participants. Each participant provided 50 utterances (on average 4.6 minutes) of training data and 20 utterances (on average 1.9 minutes) of test data. The prompts for each user covered five names, each with 10 training utterances and 4 test utterances, with each name potentially appearing multiple times per utterance. The dataset includes accented and disfluent speech. 

We used the Wiki-Names dataset for personalization experiments. The baseline ASR model is a recurrent neural network transducer (RNN-T)~\cite{graves2012sequence} as described in~\cite{he2019streaming}.
The models were trained using the efficient implementation~\cite{bagby2018efficient} in TensorFlow~\cite{abadi2016tensorflow}. We measured the success of the modified model using the word error rate (WER) metric as well as the name recall rate~\cite{sim19asru} as described below:
\begin{equation}
    \text{recall} = \frac{\text{retrieved} \cap \text{relevant}}{\text{relevant}}
\end{equation}	
where {\tt retrieved} is the number of times names are present in the hypotheses and {\tt relevant} refers to the number of times names appear in the reference transcripts.  {\tt retrieved} $\cap$ {\tt relevant} indicates the number of relevant names that are correctly retrieved.

In addition to tracking quality metrics, we also quantify the impact of the algorithms on training memory by running on-device benchmarks. Comparisons were made between different parameterization ranks, different optimizers, and full-rank, baseline model.

\subsection{Memory Benchmark}
\label{sec:memroy_benchmark}

\begin{figure}[t]
    \centering
    \includegraphics[width=0.5\textwidth]{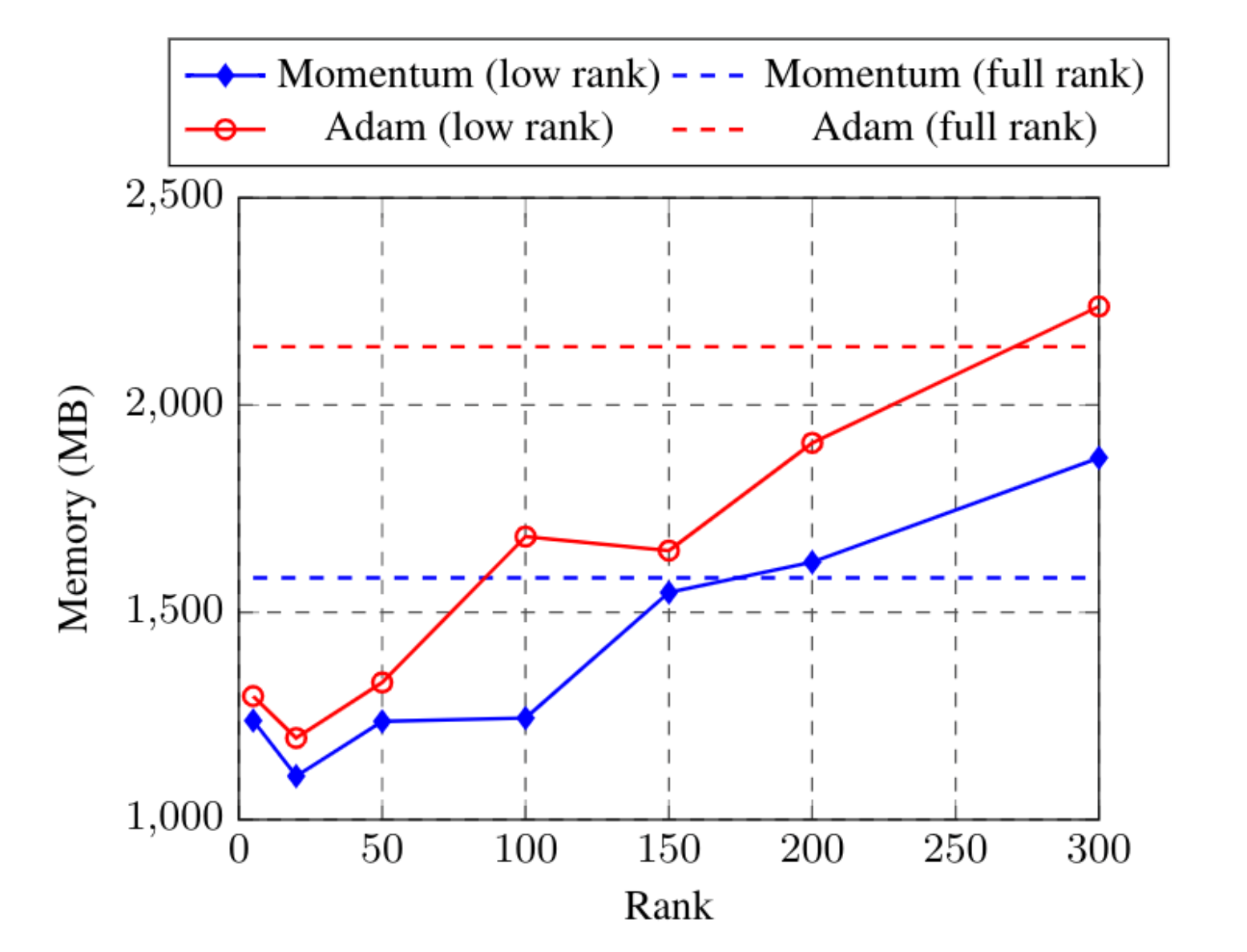}
    \caption{Comparison of peak training memory (in megabytes) used for Momentum and Adam optimizers with different ranks.}
    \label{fig:memory}
\end{figure}
The low-rank gradient model saved a significant amount of memory. Figure~\ref{fig:memory} shows the training memory with low-rank gradient projection versus the baseline (full rank) models, for both the momentum and Adam optimization methods.
We adjusted the rank of the gradient projection matrix across experiments to observe the impact on memory. 
Figure~\ref{fig:memory} shows that the low-rank model uses less memory than the full-rank model using the momentum optimizer for a projection of rank 100. Any projection of a lower rank would continue to save memory. Similarly, the modified model was able to save training memory with the Adam optimizer for a projection of up to rank 200. Additionally, the graph illustrates that the training memory increases about linearly with rank.
Furthermore, with rank 100 and 150, low-rank gradient projection with Adam optimization consumes about the same memory as full rank momentum optimization. 

\subsection{Speech Recognition Performance}
\label{sec:wer}

\begin{figure}[t]
    \centering
    \includegraphics[width=0.46\textwidth]{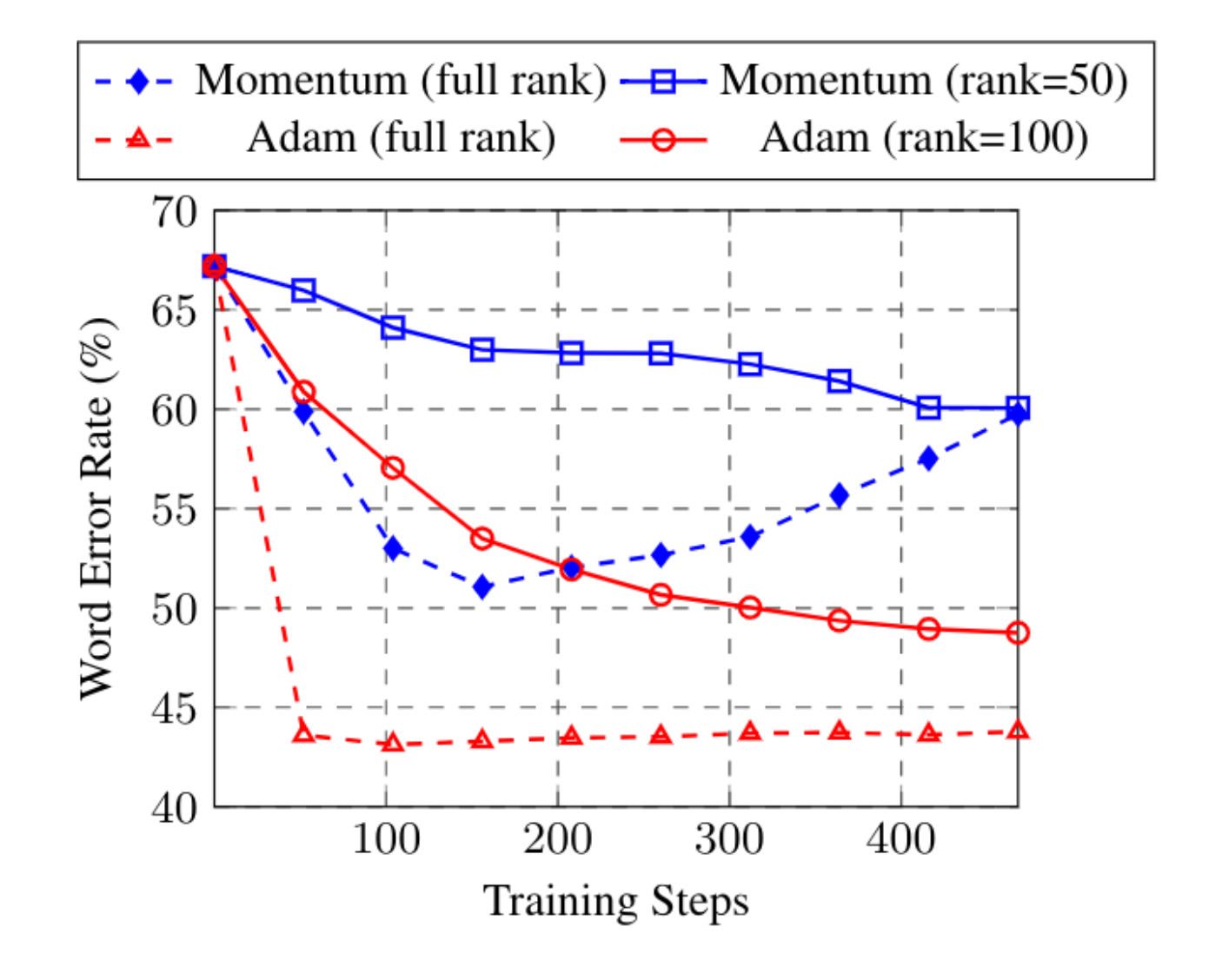}
    \caption{Comparison of word error rate performance for Momentum and Adam optimization with and without low-rank approximation.}
    \label{fig:wer_compare_optimizers}
\end{figure}
\begin{figure}[t]
    \centering
    \includegraphics[width=0.46\textwidth]{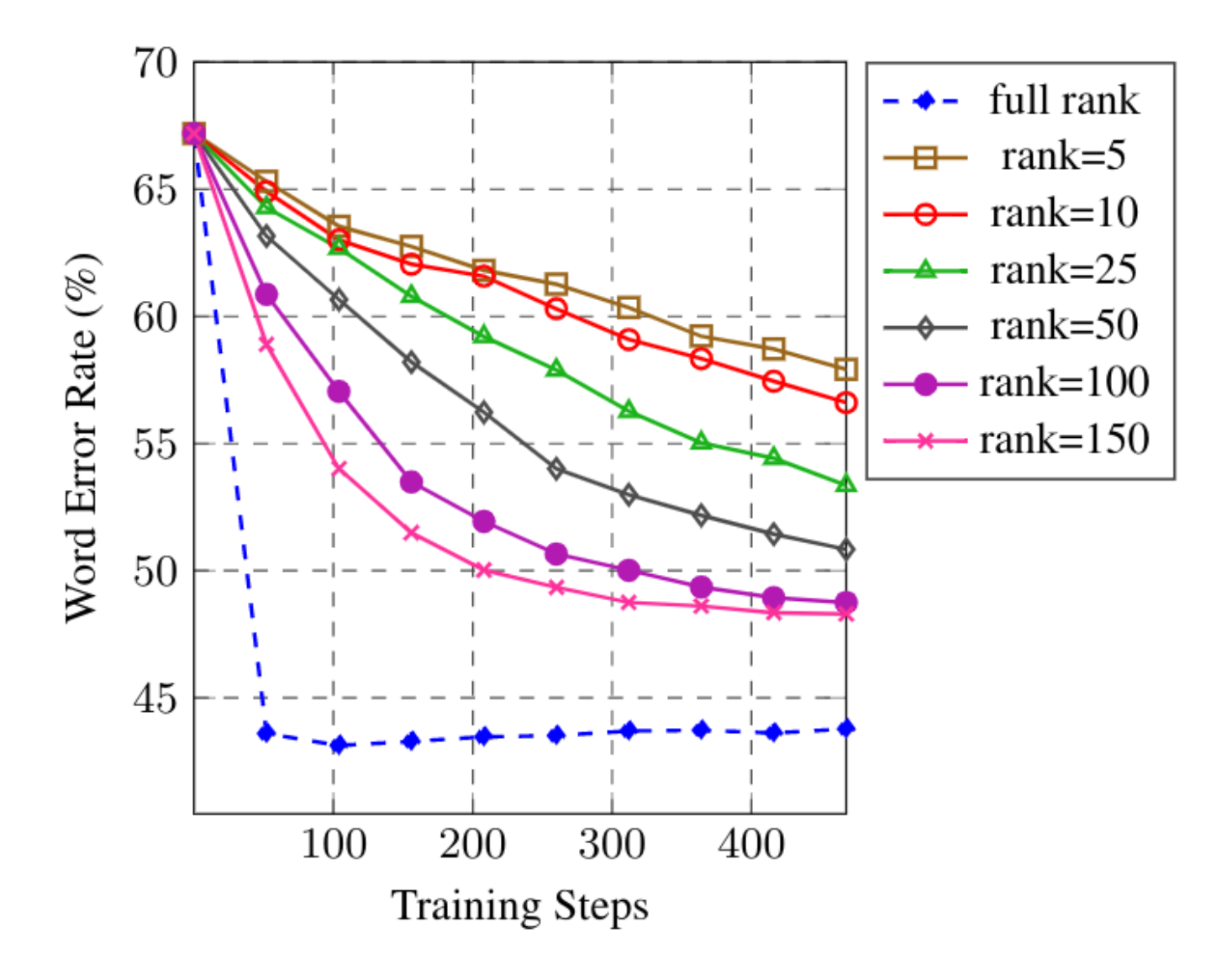}
    \caption{Comparison of word error rate performance for Adam optimization with different ranks.}
    \label{fig:wer_vs_rank}
\end{figure}

The results in Figures~\ref{fig:wer_compare_optimizers} and~\ref{fig:wer_vs_rank} show how the speech recognition quality varies with increasing training steps for different settings. 
Figure~\ref{fig:wer_compare_optimizers} compares the WER for the momentum and Adam optimization methods with and without low-rank projection. Comparing the low-rank and full-rank models, Figure~\ref{fig:wer_compare_optimizers} shows that low-rank models converge slower for both momentum and Adam optimization. The latter achieved a better performance, indicating that we are able to take advantage of the benefit of Adam optimization by using it to update $\U$ and $\V$.
Figure~\ref{fig:wer_vs_rank} shows that the word error rate decreases faster and to a lower rate as the rank of the gradient approximation is increased. Training the model using Adam with a gradient projection matrix of rank 150 reached a word error rate of 47.1\% while the baseline model converges at a word error rate about 43.8\%. 
\begin{figure}[t]
    \centering
    \includegraphics[width=0.46\textwidth]{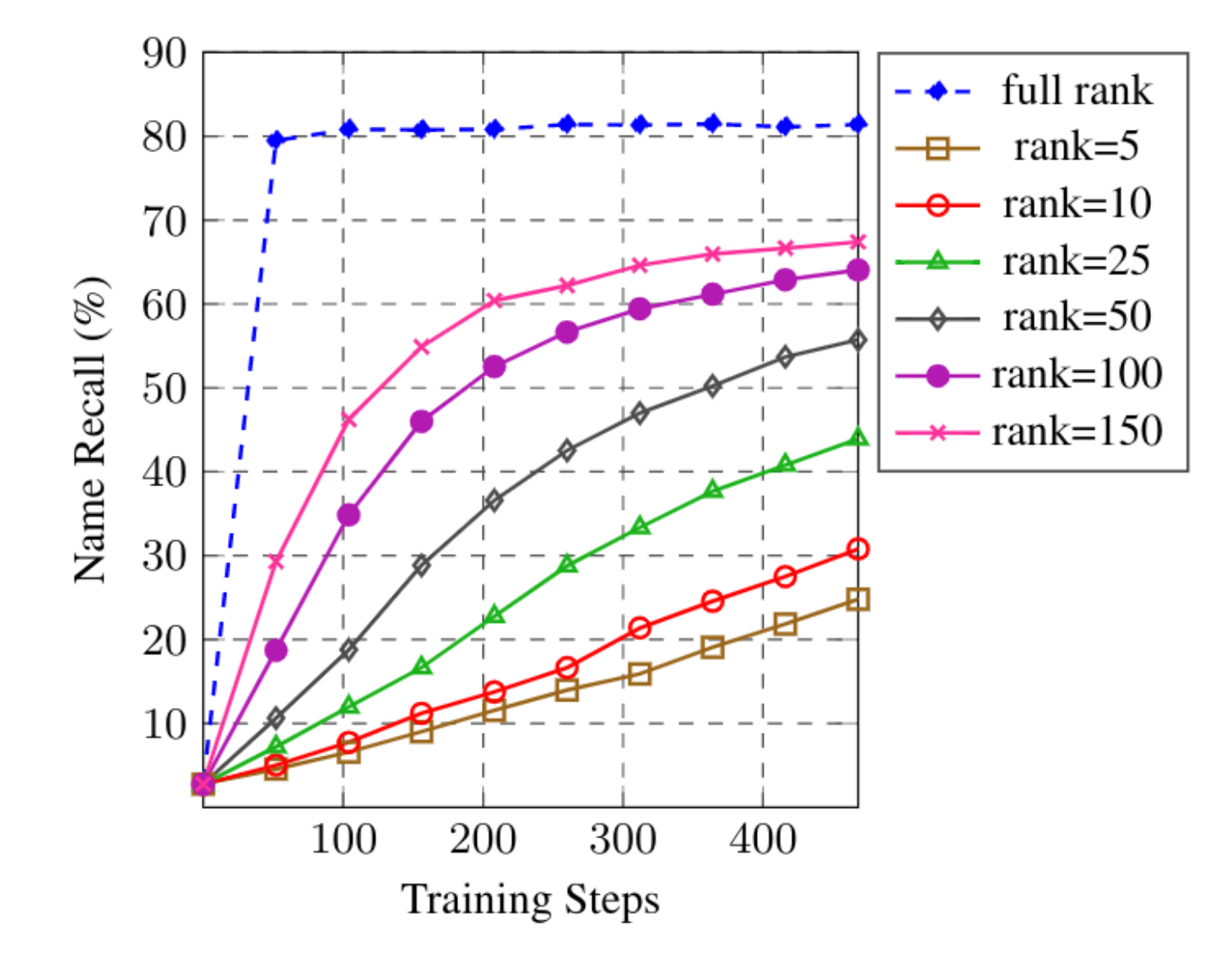}
    \caption{Comparison of name recall rate performance for Adam optimization with different ranks.}
    \label{fig:recall_vs_rank}
\end{figure}
Similarly, Figure~\ref{fig:recall_vs_rank} shows that the name recall rate increases faster and to a higher rate with the higher rank models, as expected.

\section{Summary}
\label{sec:conclusions}

The experiments detailed above sought to explore an opportunity to save training memory for deep neural network models, such as those used for speech recognition. 
Approximating the gradient computation using low-rank parameters saves memory up to a rank of about 100 and 200 for the momentum and Adam optimization, respectively. These results are promising.

To observe the impact of the new training method on the effectiveness of the model, we did experiments on Wiki-Names, a dataset with accented speech and difficult-to-recognize names. The most important metrics from these experiments are the word error rate and the recall for the names in the dataset. We compared how models of different ranks and different optimizers trained and how their training compared to the baseline model. For the model using the momentum optimizer, the rank did not impact training significantly. Furthermore, the low-rank model for momentum performed worse than the baseline momentum model. Predictably, the low-rank model with the Adam optimizer performed much better. Additionally, rank had an observable impact on training.

Using a low-rank approximation of the gradient computation for deep neural network models provides an opportunity to save memory without a significant increase in error rate or decrease in recall rate. This opportunity is most promising for on device training with more advanced optimizers, like Adam, that traditionally use multiple high-dimensional parameters for gradient computation.

\bibliographystyle{IEEEbib}
\bibliography{refs}

\begin{thebibliography}{10}

\bibitem{hinton2012deep}
Geoffrey Hinton, Li~Deng, Dong Yu, George~E Dahl, Abdel-rahman Mohamed, Navdeep
  Jaitly, Andrew Senior, Vincent Vanhoucke, Patrick Nguyen, Tara~N Sainath,
  et~al.,
\newblock ``Deep neural networks for acoustic modeling in speech recognition:
  The shared views of four research groups,''
\newblock {\em IEEE Signal Processing Magazine}, vol. 29, no. 6, pp. 82--97,
  2012.

\bibitem{he2019streaming}
Yanzhang He, Tara~N Sainath, Rohit Prabhavalkar, Ian McGraw, Raziel Alvarez,
  Ding Zhao, David Rybach, Anjuli Kannan, Yonghui Wu, Ruoming Pang, et~al.,
\newblock ``Streaming end-to-end speech recognition for mobile devices,''
\newblock in {\em IEEE International Conference on Acoustics, Speech and Signal
  Processing (ICASSP)}. IEEE, 2019, pp. 6381--6385.

\bibitem{sim19interspeech}
Khe~Chai Sim, Petr Zadrazil, and Fran{\c{c}}oise Beaufays,
\newblock ``An investigation into on-device personalization of end-to-end
  automatic speech recognition models,''
\newblock in {\em Interspeech}, 2019.

\bibitem{sim19asru}
Khe~Chai Sim, Fran{\c{c}}oise Beaufays, Arnaud Benard, Dhruv Guliani, Andreas
  Kabel, Nikhil Khare, Tamar Lucassen, Petr Zadrazil, Harry Zhang, Leif
  Johnson, Giovanni Motta, and Lillian Zhou,
\newblock ``Personalization of end-to-end speech recognition on mobile devices
  for named entities,''
\newblock {\em to appear in ASRU}, 2019.

\bibitem{xue2014singular}
Jian Xue, Jinyu Li, Dong Yu, Mike Seltzer, and Yifan Gong,
\newblock ``Singular value decomposition based low-footprint speaker adaptation
  and personalization for deep neural network,''
\newblock in {\em Proc. ICASSP}. IEEE, 2014, pp. 6359--6363.

\bibitem{zhao2016low}
Yong Zhao, Jinyu Li, and Yifan Gong,
\newblock ``Low-rank plus diagonal adaptation for deep neural networks,''
\newblock in {\em Proc. ICASSP}. IEEE, 2016, pp. 5005--5009.

\bibitem{samarakoon2016factorized}
Lahiru Samarakoon and Khe~Chai Sim,
\newblock ``Factorized hidden layer adaptation for deep neural network based
  acoustic modeling,''
\newblock {\em IEEE/ACM Transactions on Audio, Speech, and Language
  Processing}, vol. 24, no. 12, pp. 2241--2250, 2016.

\bibitem{nash1990singular}
JC~Nash,
\newblock ``The singular-value decomposition and its use to solve least-squares
  problems,''
\newblock {\em Compact Numerical Methods for Computers: Linear Algebra and
  Function Minimisation}, pp. 30--48, 1990.

\bibitem{mcclelland1987parallel}
James~L McClelland, David~E Rumelhart, PDP~Research Group, et~al.,
\newblock {\em Parallel distributed processing}, vol.~2,
\newblock MIT press Cambridge, MA:, 1987.

\bibitem{sutskever2013importance}
Ilya Sutskever, James Martens, George Dahl, and Geoffrey Hinton,
\newblock ``On the importance of initialization and momentum in deep
  learning,''
\newblock in {\em International conference on machine learning}, 2013, pp.
  1139--1147.

\bibitem{kingma2015adam}
Diederik~P. Kingma and Jimmy Ba,
\newblock ``Adam: {A} method for stochastic optimization,''
\newblock in {\em 3rd International Conference on Learning Representations,
  {ICLR} 2015, San Diego, CA, USA, May 7-9, 2015, Conference Track
  Proceedings}, 2015.

\bibitem{graves2012sequence}
Alex Graves,
\newblock ``Sequence transduction with recurrent neural networks,''
\newblock {\em arXiv preprint arXiv:1211.3711}, 2012.

\bibitem{bagby2018efficient}
Tom Bagby, Kanishka Rao, and Khe~Chai Sim,
\newblock ``Efficient implementation of recurrent neural network transducer in
  {TensorFlow},''
\newblock in {\em 2018 IEEE Spoken Language Technology Workshop (SLT)}. IEEE,
  2018, pp. 506--512.

\bibitem{abadi2016tensorflow}
Mart{\'{\i}}n Abadi, Ashish Agarwal, Paul Barham, Eugene Brevdo, Zhifeng Chen,
  Craig Citro, Gregory~S. Corrado, Andy Davis, Jeffrey Dean, Matthieu Devin,
  Sanjay Ghemawat, Ian~J. Goodfellow, Andrew Harp, Geoffrey Irving, Michael
  Isard, Yangqing Jia, Rafal J{\'{o}}zefowicz, Lukasz Kaiser, Manjunath Kudlur,
  Josh Levenberg, Dan Man{\'{e}}, Rajat Monga, Sherry Moore, Derek~Gordon
  Murray, Chris Olah, Mike Schuster, Jonathon Shlens, Benoit Steiner, Ilya
  Sutskever, Kunal Talwar, Paul~A. Tucker, Vincent Vanhoucke, Vijay Vasudevan,
  Fernanda~B. Vi{\'{e}}gas, Oriol Vinyals, Pete Warden, Martin Wattenberg,
  Martin Wicke, Yuan Yu, and Xiaoqiang Zheng,
\newblock ``Tensorflow: Large-scale machine learning on heterogeneous
  distributed systems,''
\newblock {\em CoRR}, vol. abs/1603.04467, 2016.

\end{thebibliography}

\end{document}